\documentstyle[aps,preprint,epsf]{revtex}

\newcommand{\size}{10.0truecm}

\begin{document}
\preprint{SUNY-NTG-98-63; JYFL 98-17}
\title{\bf $e^+e^-$ yields in Pb+Au collisions at 158 AGeV/c: Assessment of
baryonic contributions \\ }

\author {Pasi Huovinen$^{1,2}$ and Madappa Prakash$^1$ }
\address{
$^1$Department of Physics \& Astronomy, SUNY at Stony Brook,
Stony Brook, USA \\
$^2$Department of Physics, University of Jyv{\"a}skyl{\"a}, Finland }
\date{\today}

\maketitle

\begin{abstract}

Using a hydrodynamic approach to describe Pb+Au collisions at 158
AGeV/c, we analyze $e^+e^-$ yields from matter containing baryons in
addition to mesons.  We employ $e^+e^-$ production rates from two
independent calculations, which differ both in their input physics and
in their absolute magnitudes, especially in the mass range where
significant enhancements over expected backgrounds exist in the CERES
data. Although the presence of baryons leads to significant
enhancement of $e^+e^-$ emission relative to that from mesons-only
matter, the calculated results fall below the data in the range
$400 < M_{e^+e^-}/{\rm MeV} < 600$, by a factor of 2-3. Since the 
calculated $e^+e^-$ spectra are relatively insensitive to the equation
of state for initial conditions that fit the observed hadronic
spectra, either in-medium modifications of the $e^+e^-$ sources more
significant than so far considered, or the presence of hitherto
unidentified additional sources of $e^+e^-$ is indicated.  \\

\noindent PACS number(s): 25.70.np, 12.38.Mh, 12.40.Ew, 47.75.+f \\
\end{abstract}


Electromagnetic signals from relativistic heavy-ion reactions directly
probe the properties of the dense matter created during the collision,
since, after production, their interactions with the evolving ambient
matter are negligible~\cite{qm96}.  The CERES collaboration at CERN,
beginning with S+Au collisions at 200 AGeV/c and more recently with
Pb+Au collisions at 158 AGeV/c, have reported dielectron ($e^+e^-$)
yields that significantly exceed those from expected
backgrounds~\cite{Agakichiev95,Drees96}. For the pair invariant masses
$M_{e^+e^-} < 2$ GeV, the background is chiefly from the decays of
neutral mesons after they cease interacting with the produced matter.
For nucleus-nucleus ($AA$) collisions, these backgrounds are estimated
by extrapolating the measured yields from proton-proton ($pp$) and
proton-nucleus ($pA$) collisions. Specifically, the reported average
enhancements are $5.0\pm 0.7(\rm stat.)\pm 2.0(\rm syst.)$ in the
range $0.2 < M_{e^+e^-}/{\rm GeV} < 1.5$ for the S+Au reaction and
$3.5\pm 0.4(\rm stat.)\pm 0.9(\rm syst.)$ in the range
$0.2 < M_{e^+e^-}/{\rm GeV} < 2$ for the Pb+Au reaction (hereafter the
'95 CERES data), respectively. In both cases, the pseudo-rapidity
coverage was in the range $2.1 < \eta < 2.65$, the transverse momenta
of the $e^+$ and $e^-$ were restricted to $p_t > 200$ MeV, and only
pairs with opening angle $\Theta_{ee} > 35~{\rm mrad}$ were accepted.
The two experiments differed in that $\langle dN_{ch}/d\eta\rangle = 125$
for S+Au and $\langle dN_{ch}/d\eta\rangle = 220$ for Pb+Au. More
recently, $e^+e^-$ data from Pb+Au collisions with improved statistics
and mass resolution have become available (hereafter the '96 CERES
data)~\cite{Lenkeit98a}. To within two standard deviations, the '96
CERES data are consistent with the '95 CERES data and show similar
enhancements, albeit at a somewhat lower level. To be specific, the
average enhancement found was $2.6\pm 0.5(\rm stat.)\pm 0.6(\rm syst.)$
in the range $0.25 < M_{e^+e^-}/{\rm GeV} < 0.7$, with the enhancement
increasing for more central collisions.  For the same mass region and
for the same background evaluation, the '95 CERES data gives an average
enhancement of $3.9\pm 0.9(\rm stat.)\pm 0.9(\rm syst.)$.  These
enhancements over the expected background, particularly in the range
$0.4 < M_{e^+e^-}/{\rm GeV} < 0.6$ in both S+Au and Pb+Au reactions,
are especially intriguing, since they have the potential to distinguish
between competing theoretical scenarios concerning the in-medium
properties of vector mesons, including those of chiral restoration with
increasing density and temperature.

The theoretical analysis of $e^+e^-$ yields in $AA$ collisions
involves integrating the microscopic production rates over the
space-time history of the produced matter and implementing the
experimental cuts and resolution for an apposite comparison with the
data. Since the dynamical evolution proceeds through stages consisting
of both sub-hadronic and hadronic degrees of freedom, a knowledge of
the relevant $e^+e^-$ production rates is essential for a complete
description. In this letter, we confront both the '95 and '96 data
from Pb+Au collisions with theoretical calculations which employ a
hydrodynamic description for the space-time evolution and which use
recently calculated $e^+e^-$ emission rates in matter including
baryons. Efforts are made to achieve a simultaneous description of
both the hadronic and electromagnetic data. Special emphasis is placed
on delineating the role baryons play in enhancing $e^+e^-$ emissions
relative to emissions in matter without baryons.

The dominant process in the Quark-Gluon Plasma (QGP) phase is the
reaction $q\bar q\rightarrow e^+e^-$, for which we use the lowest order
(in $\alpha$ and $\alpha_s$) rates at a finite baryon chemical
potential, summarized in Ref.~\cite{Cleymans87}. Rates of
${\cal O}(\alpha^2\alpha_s)$ have significant contributions only in the
mass region $M_{e^+e^-} < 500$ MeV, where the Dalitz decays of the final
state mesons dominate, and are therefore not important.  In all of the
results to be reported, contributions from $q\bar q$ processes, whether
they are from a pure QGP phase or a mixed phase, are dominated by those
arising from the hadron gas phase.

In the hadron gas phase, prompt decays from vector, pseudoscalar, and
axial vector mesons, as well as radiative decays and bremsstrahlung
processes involving these mesons, all yield significant $e^+e^-$
contributions in the invariant mass regions of interest
here~\cite{Gale91,Gale94,Lichard94,Gale98}.
Through independent calculations, we have verified that the rates
calculated by Gale and Lichard~\cite{Gale94} (GL hereafter) and
Lichard~\cite{Lichard94}, who delineate a number of these processes,
are consistent with the spectral-function approach of Steele,
Yamagishi, and Zahed~\cite{Steele97} (SYZ hereafter), who use
experimentally extracted spectral functions and on-shell chiral
reduction formulas coupled with a virial expansion scheme. Thus,
in mesons-only matter, the GL and SYZ rates can be taken as
a standard against which comparisons with other
approaches~\cite{Chan96,Rapp97,Rapp98,Rapp98c,Klingl96,Song96,Friman97,Peters98,Eletsky98}, in which
possible additional in-medium many-body effects are considered, may be
made. These ``standard'' rates, when convolved  with the space-time
evolution of matter, in both transport and hydrodynamic
approaches~\cite{Li95,Cassing95,Cassing98,Sri96,Sollfrank97a,Hung97},
have yielded results which are significantly below the measured yields
in the region $400 < M_{e^+e^-}/{\rm MeV} < 600$ in both S+Au and P+Au
collisions. To date, only transport models~\cite{Li95,Cassing95,Cassing98},
which incorporate substantial modifications of the in-medium
properties, such as a sharply decreasing in-medium $\rho$-meson
mass~\cite{Brown91} or a significant widening of its
width~\cite{Chan96}, have been able to account for the  data.

In order to examine the extent to which baryons influence $e^+e^-$
yields, we utilize rates from the independent calculations of Rapp,
Urban, Buballa, and Wambach~\cite{Rapp98} (RUBW hereafter), and
Steele, Yamagishi, and Zahed~\cite{Steele97}. We have chosen these
two rates, both because they represent a contrast in terms of the
input physics  and theoretical techniques employed, and because they
differ significantly in the absolute magnitudes, especially for
$M_{e^+e^-} \approx 500$ MeV, where the discrepancy between
calculations and data is  most significant.

The RUBW rates are based on a many-body approach in which
phenomenological interactions are used to calculate the $\rho$-meson
spectral function in matter containing baryons. In the first
attempts using this approach~\cite{Chan96,Rapp97}, nucleon-hole and
delta-hole excitations were found to substantially modify the
$\rho$-meson spectral function from its form in vacuum and to
significantly enhance the $e^+e^-$ yields relative to those without
such modifications. In subsequent works~\cite{Rapp98,Rapp98c},
however, these medium effects were constrained by photoabsorption data
on protons and nuclei and on $\pi N \rightarrow \rho N$ production
experiments. $\rho N$ interactions, particularly s-wave interactions
involving the resonance N$^*(1520)$ and, to a lesser extent, p-wave
interactions with the N$^*(1720)$, were found to result in a
substantial broadening of the in-medium $\rho$-meson spectral
function, albeit at a somewhat lower level than found in earlier
works~\cite{Chan96,Rapp97}. A theoretical discussion of the
similarities and differences between the models of RUBW and those
based on the Brown-Rho scaling of vector meson masses~\cite{Brown91}
may be found in Ref.~\cite{Brown98}.

Independently, SYZ have extended the spectral-function approach to
include baryons~\cite{Steele97}. To leading order in the baryon
density, the dilepton rates are proportional to the spin-averaged
forward Compton scattering amplitude on the baryons by off-shell
photons with $q^2\geq 0$. While the photon rates can be determined
directly from data by use of the optical theorem, the dilepton rates
require additional theoretical considerations. SYZ use chiral
constraints to determine the tree-level strong-interaction Lagrangian
and preserve perturbative unitarity from an on-shell loop-expansion
in $1/f_\pi$, which enforces current conservation and crossing
symmetry. To one-loop, the rates are calculated without the use of any
free parameter. The large contribution of the $\Delta$ to the Compton
amplitude near threshold is taken into account by adding it as a
unitarized tree term to the one-loop result. SYZ find that the
dominant contribution arises from pion-nucleon interactions in the
continuum and not from the $\Delta$ resonance.

\begin{center}
  \begin{figure}
    \hspace*{-10mm}
     \epsfxsize \size \epsfbox{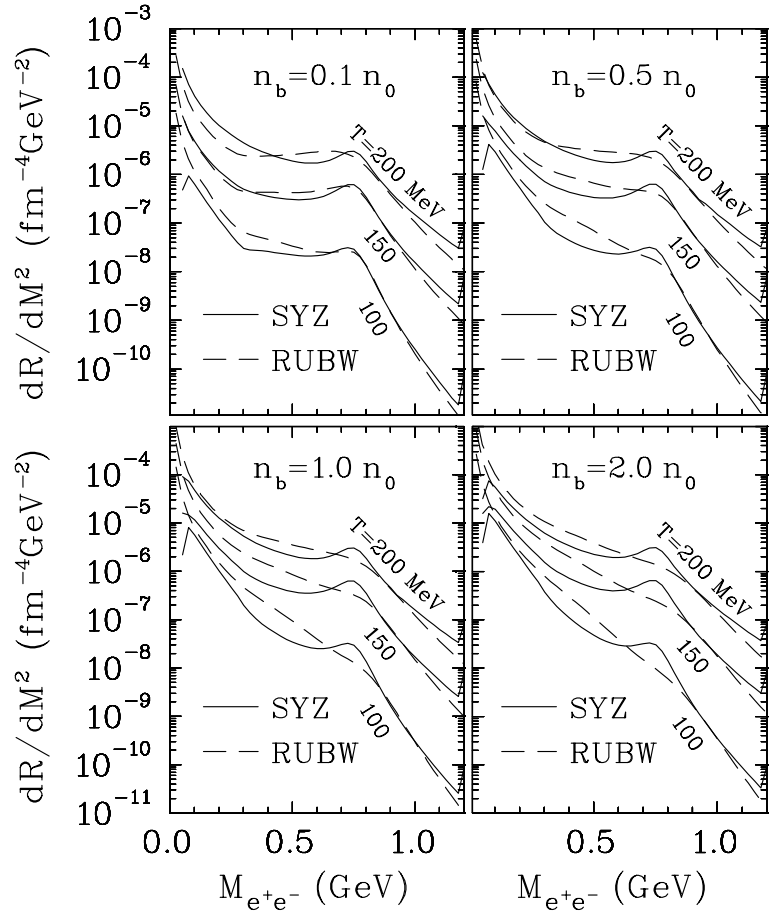}
     \caption{The $e^+e^-$ production rates at different temperatures and
		baryon densities versus pair invariant mass. The solid
		lines are results of Steele, Yamagishi, and Zahed (SYZ)
		and the dashed lines are those of Rapp, Urban, Buballa,
		and Wambach (RUBW).}
     \label{rates}
  \end{figure}
\end{center}

The SYZ and RUBW rates are compared in Fig.~\ref{rates}, which
illustrates the extent to which baryons can enhance the rates relative
to those in mesons-only matter. For typical values of the temperature
and baryon density of relevance here, the distinguishing features of
the SYZ rates are: (1) Enhancements relative to the baryon-free case
are of order 2-3 and are restricted to
$M_{e^+e^-}/{\rm MeV} < 500$. (2) The prominent signature at the
$\rho$-meson vacuum mass persists at nearly all values of $T$ and
$n_b$. In contrast to SYZ, the two most striking features of the RUBW
results are: (1) Rates are significantly larger than those of SYZ in
the range $ 200 < M_{e^+e^-}/{\rm MeV} < 600$. (2) The tell-tale
signature at the $\rho$-meson vacuum mass is weakened, predominantly
with increasing $n_b$. The differences between the results of these
two calculations underscore the fact that a standard calculation of
the rates in matter containing baryons, similar to those in
mesons-only matter, is not yet available, which stresses the need to
identify both common and distinguishing features between these two
approaches.

As in Ref.~\cite{Sollfrank97a}, where a hydrodynamical description of
200 AGeV/c S+Au collisions was given, we place emphasis on the
simultaneous description of the hadron and electromagnetic data in 158
AGeV/c Pb+Au collisions. Details of the hydrodynamical calculations
may be found in Refs.~\cite{Sollfrank97a,Huovinen98}. Following these
works, we assume that locally the system rapidly achieves thermal and
chemical equilibrium and parameterize it in terms of the various
particle  densities and fluid velocity. Our parametrization of the
initial baryon density profile is motivated by the gross features of
$pp$ spectra (see Refs.~\cite{Sollfrank97b,Sollfrank97c}). The values
of parameters describing the initial conditions are chosen to
reproduce the measured hadron spectra. The densities and
temperatures for the various stages of evolution are determined by
specifying the equation of state (EOS), which enables a prediction of
the $e^+e^-$ emission from the various stages of the collision.

In view of the fact that hadronic data for the Pb+Au collision are not
yet available, we have used the NA49 data from the Pb+Pb collision
\cite{Jones96} to parameterize the initial conditions. The difference
between the sizes of Pb and Au nuclei is small, and therefore one may
justifiably expect the differences between the hadronic spectra in
these two cases to be negligible. The initial conditions and the
resulting hadronic spectra in Pb+Pb collisions are detailed in
Ref.~\cite{Huovinen98}. It must be noted, however, that NA49 uses
a different centrality trigger from that used by CERES. Our
calculations, which are tuned to reproduce the results of NA49, yield
an average multiplicity of $\langle dN_{ch}/d\eta\rangle \cong 330$
within the CERES acceptance region. The CERES collaboration finds both
the shape of the spectrum and the yield scaled with multiplicity to
vary with multiplicity~\cite{Ravinovich98}. We have therefore opted
here to compare our calculated results with the preliminary
data~\cite{Voigt98,Lenkeit98b} from nearly central collisions with
$\langle dN_{ch}/d\eta\rangle = 350$. We wish to emphasize that these
unpublished data, in contradistinction to the '95 and '96 CERES data,
await confirmation from the CERES collaboration\footnote{We acknowledge
	communications from J.~Stachel and I.~Tserruya, who
	alerted us to the preliminary nature of these data
	and to the ongoing intense efforts of the CERES
	collaboration to release confirmed data.}.
In the text and figures that follow,  we refer to the data from Voigt's
thesis~\cite{Voigt98} as '95 data and that from Lenkeit's
thesis~\cite{Lenkeit98b} as '96 data. Comparisons of calculated results
with the CERES data for noncentral collisions with different
multiplicities and for different dielectron $p_t$ cuts are essential
for a comprehensive understanding, and will be reported separately.

For the most part, we show results of a baseline model calculated
using the EOS labeled EOS D and initial state labeled IS 1 in
Ref.~\cite{Huovinen98}. For this EOS, the phase transition from a
hadron gas to the QGP occurs at a temperature $T_c= 200$ MeV. The main
reason for using an EOS with the somewhat high $T_c$ of 200 MeV was to
prolong the lifetime of the hadronic phase and make the changes in the
emission rates due to the presence of baryons as distinct as possible
from the case in which their contributions are ignored. Comparisons
with the case in which $T_c=165$ MeV (EOS A and IS 1 in
Ref.~\cite{Huovinen98}) will highlight the dependence on $T_c$.

\begin{center}
  \begin{figure}
    \hspace*{-10mm}
     \epsfxsize \size \epsfbox{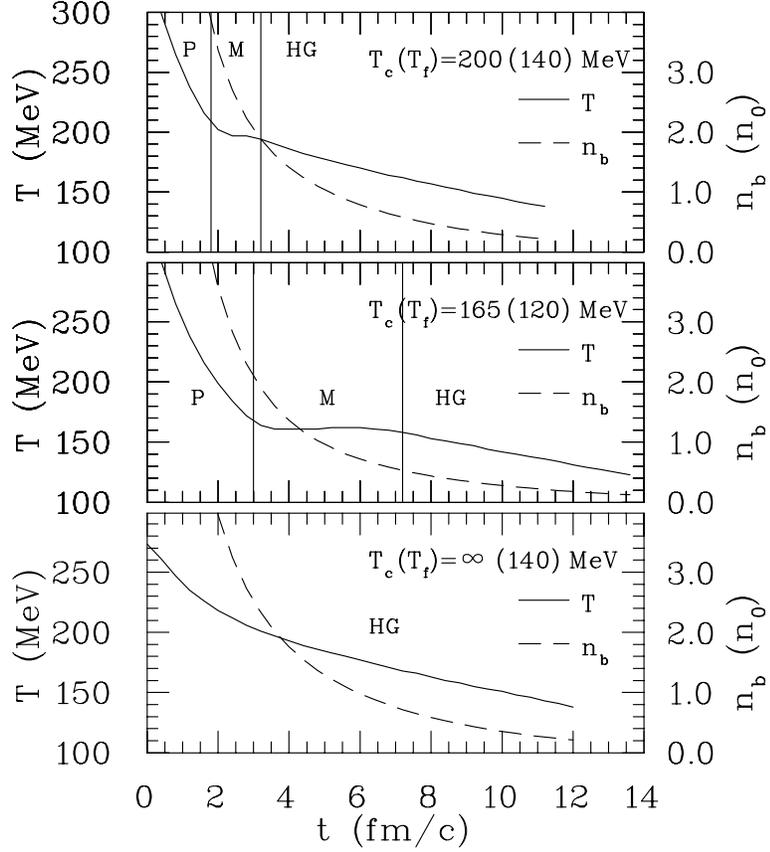}
     \caption{Temperature (left scale) and baryon density in units of the
		nuclear equilibrium density (right scale) in the center of
		the fireball as a function of time. The symbols P, M, and
		HG denote the plasma, the mixed, and hadron gas phases,
		respectively. The top (middle) panel shows results for
		an EOS with $T_c=200(165)$ MeV and the bottom panel for
		a hadronic EOS without a phase transition. Equivalent fits
		to the hadronic spectra are obtained at the indicated
		freeze-out temperatures of $T_f=120$ and 140 MeV.}
     \label{evol}
  \end{figure}
\end{center}

The hydrodynamical approach requires us to define a physical stage at
which the system is sparse enough to be treated as noninteracting and
consisting of freely streaming particles. For our baseline model, we
define this so-called freeze-out as a space-time surface of constant
energy density of $\epsilon_f = 0.15(0.069)$ GeV/fm$^3$, which leads
to an average freeze-out temperature of $T_{f} \cong 140(120)$ MeV.
In~Ref.~\cite{Huovinen98},  equivalent best fits to the hadron data
were obtained using an EOS with $T_c = 200$ MeV combined with
$T_f=140$ MeV, and also an EOS with  $T_c = 165$ MeV with $T_f = 120$
MeV. In the top and middle panels of Fig.~\ref{evol}, we compare the
evolution of the temperature and baryon number density at the center
of the fireball for these two cases. Note that, although the times for
which the hadron gas phase lasts are nearly the same in both cases,
its temperature is noticeably lower in the case in which $T_c$ and
$T_f$ are smaller. As our results will show, this will have important
implications for the total $e^+e^-$ spectrum, which is calculated by
integrating the emission rate in unit volume over the total space-time
volume enclosed by the freeze-out surface.

\begin{center}
  \begin{figure}
    \hspace*{-10mm}
     \epsfxsize \size \epsfbox{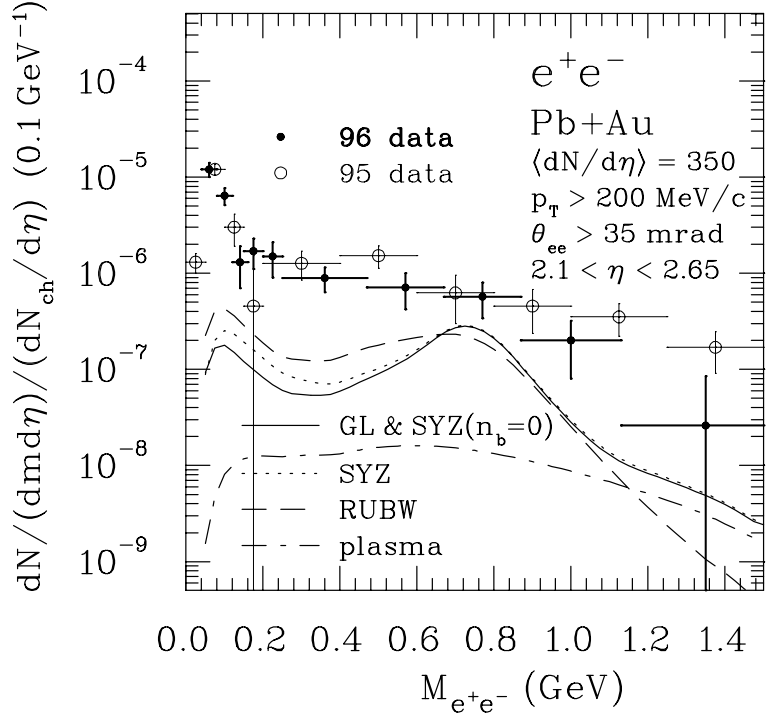}
     \caption{Dielectron yields from the hadronic part and plasma part
              of the fireball for $T_c(T_f)=200(140)$ MeV. Solid lines
              are results for matter without baryons and are obtained
              by using the rates of SYZ which agree with those of Gale
              and Lichard (GL). Results in matter with baryons
              (short-dashed lines from the SYZ rates and long-dashed
              lines from the rates of RUBW) are also shown.
              The data shown are preliminary and are from the
              theses of Voigt~\protect\cite{Voigt98} ('95 data) and
              Lenkeit~\protect\cite{Lenkeit98b} ('96 data), respectively.
              Kinematic cuts and detector resolution are incorporated.}
    \label{thermal}
  \end{figure}
\end{center}

In Fig.~\ref{thermal}, we show the dileptons radiated during the
lifetime of the fireball for the case in which $T_c(T_f)=200(140)$
MeV. These results are folded with the cuts and resolution of CERES
(for details, see Ref.~\cite{Sollfrank97a}). The role of the baryons
may be gauged by inspecting the time evolution of the baryon density
in the fireball (Fig.~\ref{evol}). In the mixed phase, the average
baryon density is $n_b = 0.4$ fm$^{-3}$; i.e.,\ 2.5 times the nuclear
equilibrium density $n_0\cong 0.16~{\rm fm}^{-3}$. However, the mixed
phase lasts for a relatively short time and occupies too small a
volume for a significant contribution to build up. Thus, in spite of
the high temperatures in the mixed phase and the enhanced rates of
both SYZ and RUBW with baryons, the mixed phase contributions to the
total yield are small.  The subleading contributions of the mixed
phase cannot be used to discriminate between competing scenarios for
the role of baryons.

The influence of baryons on $e^+e^-$ production is dominant in
thehadron gas phase, which dilutes rapidly due to hydrodynamic
expansion. For $T_c=200(165)$ MeV and $T_f=140(120)$ MeV, the
resulting average baryon density in the hadron gas phase of
$n_b = 0.08(0.03)$ fm$^{-3}$; i.e.,\ 0.5($\sim 0.2$)~$n_0$.
The SYZ rates including baryons are about a factor of $2$ larger
than those without baryons, but mostly below $M_{e^+e^-}=400$ MeV.
This translates to an enhancement of about a factor of two or less
in this region, relative to the baryon-free case. The larger rates
of RUBW result in enhancements of the thermal yield, even up to
$M_{e^+e^-}=300-600$ MeV, by a factor of about three relative to
those in mesons-only matter.

\begin{center}
  \begin{figure}
    \hspace*{-10mm}
     \epsfxsize \size \epsfbox{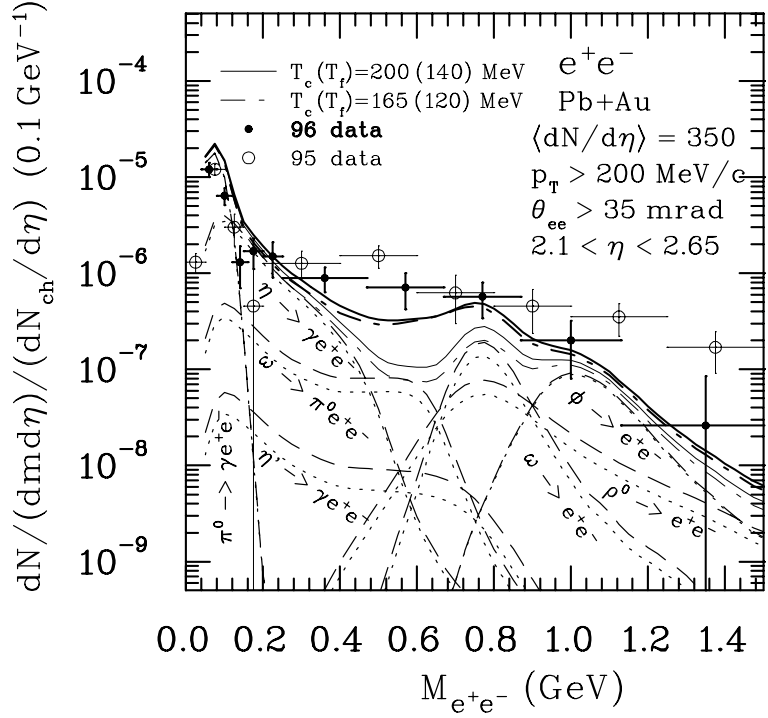}
     \caption{Background contributions from meson decays after freeze-out
		of the hydrodynamical evolution for $T_c(T_f)=200(140)$ MeV
		and 165(120) MeV, respectively, which give equivalent fits
		to the hadronic spectra. In all cases, the higher yields
		refer to $T_c(T_f)=200(140)$ MeV. The thin solid (dashed)
		line shows the total background. The thick solid (dashed)
		line shows the sum of background and fireball contributions
		calculated using the RUBW rates. The data shown are
		preliminary and are from the theses of 
		Voigt~\protect\cite{Voigt98} ('95 data) and
		Lenkeit~\protect\cite{Lenkeit98b} ('96 data),
		respectively. Kinematic cuts and detector
              resolution are incorporated.}
     \label{back}
  \end{figure}
\end{center}

In addition to $e^+e^-$ pairs from the fireball, the measured yields
contain contributions from vector meson decays after freeze-out. This
background was calculated from the distributions of hadrons at
freeze-out predicted by our model. Our calculated background is
basically in agreement with the estimated CERES background and the
experimentally measured hadron yields. The only exception is the
$\phi/h^-$ ratio, which in experiments~\cite{Friese97} is reported to
be somewhat  smaller than the ratio predicted in our calculations for
$T_f = 140$ MeV. To achieve consistency with the data, we have
suppressed the $\phi$-yield in this case by a factor of 0.6. All other
particle yields were assumed to arise from a thermally and chemically
equilibrated (in the local sense) hadron gas.  Fig.~\ref{back} shows a
comparison of the individual contributions from the various hadrons
after freeze-out for $T_c(T_f)=200(140)$ MeV (dashed curves), and
$T_c(T_f)=165(120)$ MeV (dotted curves), respectively. These results
also incorporate the experimental cuts and resolution of CERES. A
notable feature in this comparison is the extent to which
contributions from the $\omega \rightarrow e^+e^-$ decays are
sensitive to the freeze-out temperatures. This is also reflected in
the differences between the total contributions from the background at
$T_f=140$ MeV (thin solid curves) and $T_f=120$ MeV (thin dashed
curves) in the $\omega$-mass region. We will return to the
significance of this dependence later.

In Fig.~\ref{back}, we also show the sum of the background and thermal
(calculated with the RUBW rates for this figure) yields along with the
data. The solid (dashed) curves show calculated results for
$T_c=200(165)$ MeV and $T_f=140(120)$ MeV. The compensating
contributions of the fireball and background components are evident in
this comparison, inasmuch as their sums are nearly identical. A
longer lifetime increases the thermal yield, but the background
originating from a thermally and chemically equilibrated source
diminishes with increasing lifetime. These compensating effects lead
to the result that the total spectra for the two cases are similar,
differing at most by a factor of 1.2 in the region where the
discrepancy between theory and experiment is significant. This
highlights the fact that admissible changes in the phase transition
and freeze-out temperatures play only a minor role in explaining the
excess around $M_{e^+e^-} \cong$ 500 MeV, but the precise values of
the microscopic rates play a more significant role. One should bear in
mind that both $T_c(T_f)=200(140)$ MeV and $T_c(T_f) = 165(120)$ MeV
produce equally good fits to the hadronic data. A more detailed
discussion of the compensating effects of freeze-out and phase
transition temperatures can be found in~\cite{Huovinen98}.

\begin{center}
  \begin{figure}
   \hspace*{-10mm}
    \epsfxsize \size \epsfbox{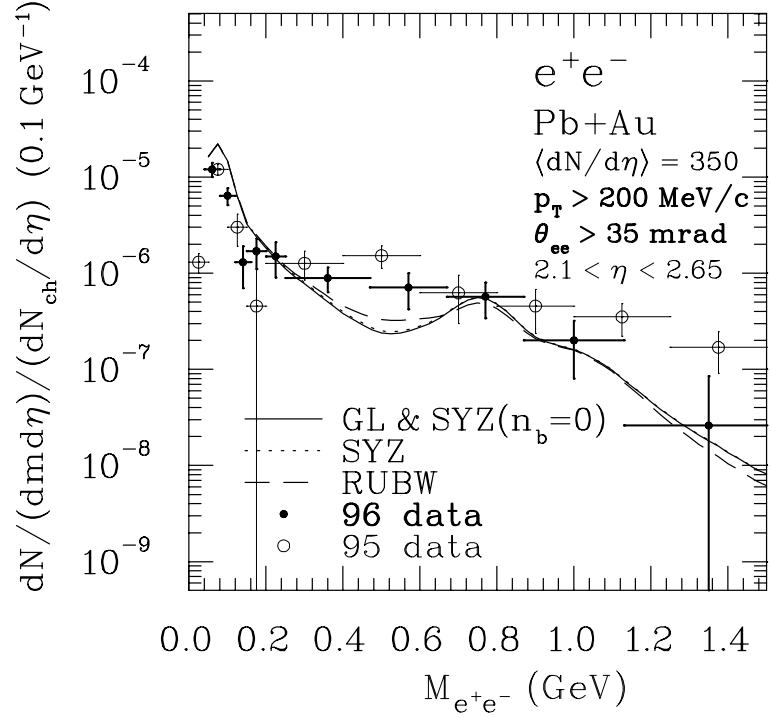}
     \caption{Calculated total dielectron mass spectra
              for $T_c(T_f)=200(140)$ MeV compared with the data.
              The data shown are preliminary and are from the theses of
	      Voigt~\protect\cite{Voigt98} ('95 data) and
              Lenkeit~\protect\cite{Lenkeit98b} ('96 data), respectively.
              Solid lines are results for matter without baryons and are
              obtained by using the rates of SYZ which agree with those
              of GL. Results in matter with baryons (short-dashed lines
              from the SYZ rates and long-dashed lines from the rates of
              RUBW) are also shown. Kinematic cuts and detector
              resolution are incorporated.}
     \label{total}
  \end{figure}
\end{center}

In Fig.~\ref{total}, a comparison of the total $e^+e^-$ mass spectrum
calculated with the rates of SYZ and RUBW with the data is shown.
Results for the SYZ rates with and without baryons are virtually
indistinguishable from each other, despite the fact that significant
enhancements were found below $M_{e^+e^-} < 400$ MeV. However, in this
mass region, the Dalitz decay backgrounds, particularly those from
$\eta \rightarrow \gamma e^+e^-$ and $\omega \rightarrow \pi^0e^+e^-$,
are an order of magnitude larger than the thermal yields and entirely
mask the baryonic contributions. Thus, the differences in the total
spectra are negligible for matter with and without baryons.  The RUBW
rates, being larger than those of SYZ in the region below the
$\rho$-mass, lead to yields that are distinguishable from the case
without baryons.  It is clear from this comparison that accounting for
the data necessitates either substantial in-medium modifications, of
magnitudes larger than so far considered, of the $e^+e^-$ sources, or
the presence of hitherto unidentified additional sources. The
theoretical justification of such strong modifications or the
existence of additional sources remains an open task.

We turn now to select comparisons with earlier works, in which the
RUBW and SYZ rates have been used to confront the data. Based on
a schematic model for both the geometry and evolution of the fireball
comprised of only hadrons from the beginning to the end of its
evolution, it has been reported in Refs.~\cite{Rapp97,Rapp98c} that
the CERES data can be accounted for by using the rates reported
there. Some relevant global features of these calculations are an
initial temperature of 170 MeV, $T_f\cong 122$ MeV, and a lifetime of
15 fm/c~\cite{Rapp98c}. In addition, a pion chemical potential was
also employed at freeze-out to reproduce the observed total
pion-to-baryon ratio.  To make an appropriate comparison, we performed
a calculation using the EOS labeled EOS H, in which only hadronic
degrees of freedom were admitted, with the initial conditions labeled
IS 1 in Ref.~\cite{Huovinen98}.  To achieve good fits to hadronic
data, an initial temperature of 270 MeV and a freeze-out temperature
of 140 MeV were required. In this case, the lifetime of the center of
the fireball was 12 fm/c~(see the bottom panel of Fig.~\ref{evol}).
The resulting $e^+e^-$ spectra are virtually identical to those shown
in Fig.~\ref{total} for the EOS with $T_c = 200$ MeV. For both EOSs
with $T_c = 200$ and 165 MeV, the system evolves through the QGP,
mixed, and pure hadron gas phases. For $T_c(T_f) = 200(140)$ MeV and
165(120) MeV, the total lifetimes of the center of the fireball are
$\cong 11(14)$~fm/c, and the lifetimes in the pure hadron gas phase
are $\cong 8(6.5)$~fm/c. In view of the inherent differences between
hydrodynamic and schematic models, particularly those related with
local versus global thermal equilibrium, chemical equilibration, and
flow effects, a one-to-one comparison of these global quantities with
those of the schematic model may be misleading.  We wish to note,
however, that in hydrodynamic calculations, the observed rapidity and
transverse momentum spectra of the hadrons are also reproduced, in
addition to the total pion-to-baryon ratio.

Our results are in qualitative agreement with those of SYZ, who also
employed a schematic model for the expanding matter and concluded that
the contributions from baryons cannot fully account for the reported
excess in the data.  This agreement, however, is mostly due to the
magnitudes of the microscopic rates, which are considerably smaller
than those of RUBW, particularly around $M_{e^+e^-}=500$ MeV. The main
conclusion that emerges from our calculations is that the $e^+e^-$
spectra are relatively insensitive to the EOS for initial conditions
that fit the observed hadronic spectra. In all cases considered here,
the calculated results fall below the data around $M_{e^+e^-}=500$ MeV
by a factor of about 2-3.

To summarize, we have calculated $e^+e^-$ emission in Pb+Au collisions
at 158 AGeV/c using two different dielectron production rates within
the framework of hydrodynamics. The rates calculated by SYZ include
baryonic contributions arising from pion-nucleon interactions and
those of RUBW account for additional in-medium modifications, which
leads to a substantial broadening of the $\rho$-meson spectral
function.  We found that the additional contributions due to baryons
in the rates of SYZ give modest contributions, but mainly at low
values of invariant mass where the spectrum is dominated by background
decays. The final dielectron spectra with and without baryonic
contributions are thus almost identical. On the other hand, the
larger $\rho$-width in the rates of RUBW leads to comparatively larger
yields in the 300--600 MeV mass region, but the enhancement still
falls short of the enhancement observed in the data.

Several important theoretical issues remain to be resolved. The most
prominent of these are: (1) The elementarity of a resonance when its
in-medium width becomes substantially larger than its vacuum width,
as found, for example, in the approach of RUBW. In many similar
instances, multi-particle multi-hole excitations have been found
necessary to account for the data. Closely connected with this is the
extent to which collectivity, as implied in the resummations employed
by RUBW, or as advocated in Ref.~\cite{Brown98}, is retained in high
energy heavy-ion collisions in which the motion of all particles is
highly randomized.  (2) A comparison of our results with those of a
transport approach to the space-time evolution of matter, in which
equivalent, if not better, fits to the data have been reported, raises
the issues of (i) whether a proper treatment of substantially
broadened resonances in a transport description has been achieved, and
(ii) whether the assumption of local chemical equilibrium in a
hydrodynamic approach is adequate to treat the observables in high
energy heavy-ion collisions.

Our calculations also highlight the need to pin down $e^+e^-$
contributions from vector meson decays after freeze-out solely from
experiments, independently of theory.  Evidently, $e^+e^-$ yields from
the decay of the $\omega$- and $\phi$-mesons are sensitive to the
freeze-out temperature in hydrodynamic calculations. In a sequential
scattering approach, the $\omega$ and $\phi$ yields will depend upon
the efficacy with which $\omega N$ and $\phi N$ absorption processes
deplete their primordial abundances~\cite{Beren94}. Clearly, data with
an improved mass resolution in the mass region of the $\omega$- and
$\phi$-mesons, especially at different bombarding energies and for
different projectile and target masses, are sorely needed to
distinguish between competing theoretical scenarios. Since the mass
shifts of these particles are expected to be small~\cite{Shuryak92},
accurate data in this mass region will be of great advantage in
calibrating the different models of space-time evolution. The
establishment of such a standard candle will be particularly important
in upcoming RHIC experiments, as it enables the disentanglement of
hadronic versus sub-hadronic contributions to be placed on a firm
footing.

We are particularly thankful to numerous intense discussions with
G.E. Brown and other members and visitors of the Nuclear Theory Group
at SUNY, Stony Brook. We gratefully acknowledge the help provided by
R. Rapp, J. Steele, and I. Zahed concerning dilepton rates in the
presence of baryons. Special thanks are due to P.V. Ruuskanen and
J. Sollfrank for a careful reading of the paper, and to
Axel Drees who provided us with the preliminary '96 data
and helped us to put the '95 data in perspective.
P.H.'s work is supported by the Academy of Finland grant 27574.
The research of M.P. is supported by the grant DOE-FG02-88ER-40388.

\bigskip



\newcommand{\IJMPA}[3]{{ Int.~J.~Mod.~Phys.} {\bf A#1}, #3 (#2)}
\newcommand{\JPG}[3]{{ J.~Phys. G} {\bf {#1}}, #3 (#2)}
\newcommand{\AP}[3]{{ Ann.~Phys. (NY)} {\bf {#1}}, #3 (#2)}
\newcommand{\NPA}[3]{{ Nucl.~Phys.} {\bf A{#1}}, #3 (#2)}
\newcommand{\NPB}[3]{{ Nucl.~Phys.} {\bf B{#1}}, #3 (#2)}
\newcommand{\PLB}[3]{{ Phys.~Lett.} {\bf {#1}B}, #3 (#2)}
\newcommand{\PRv}[3]{{ Phys.~Rev.} {\bf {#1}}, #3 (#2)}
\newcommand{\PRC}[3]{{ Phys.~Rev. C} {\bf {#1}}, #3 (#2)}
\newcommand{\PRD}[3]{{ Phys.~Rev. D} {\bf {#1}}, #3 (#2)}
\newcommand{\PRL}[3]{{ Phys.~Rev.~Lett.} {\bf {#1}}, #3 (#2)}
\newcommand{\PR}[3]{{ Phys.~Rep.} {\bf {#1}}, #3 (#2)}
\newcommand{\ZPC}[3]{{ Z.~Phys. C} {\bf {#1}}, #3 (#2)}
\newcommand{\ZPA}[3]{{ Z.~Phys. A} {\bf {#1}}, #3 (#2)}
\newcommand{\JCP}[3]{{ J.~Comp.~Phys.} {\bf {#1}}, #3 (#2)}
\newcommand{\HIP}[3]{{ Heavy Ion Physics} {\bf {#1}}, #3 (#2)}
\newcommand{\EPC}[3]{{ Eur.~Phys.~J.~C} {\bf {#1}}, #3 (#2)}

{}


\end{document}